\input harvmac
\input graphicx
\input color

\def\Title#1#2{\rightline{#1}\ifx\answ\bigans\nopagenumbers\pageno0\vskip1in
\else\pageno1\vskip.8in\fi \centerline{\titlefont #2}\vskip .5in}

%
%
\ifx\includegraphics\UnDeFiNeD\message{(NO graphicx.tex, FIGURES WILL BE IGNORED)}
\def\figin#1{\vskip2in}
\else\message{(FIGURES WILL BE INCLUDED)}\def\figin#1{#1}
\fi
\def\Fig#1{Fig.~\the\figno\xdef#1{Fig.~\the\figno}\global\advance\figno
 by1}
%
%
%
%
\def\Ifig#1#2#3#4{
\goodbreak\midinsert
\figin{\centerline{
\includegraphics[width=#4truein]{#3}}}
\narrower\narrower\noindent{\footnotefont
{\bf #1:}  #2\par}
\endinsert
}

\font\ticp=cmcsc10

\def \purge#1 {\textcolor{magenta}{#1}}
\def \new#1 {\textcolor{blue}{#1}}

\def \lvert {|}
\def \rvert {|}
\def\\{\cr}
\def\text#1{{\rm #1}}
\def\frac#1#2{{#1\over#2}}

\def \lket {|}
\def \rket {\rangle}
\def \lbra {\langle}
\def \rbra {|}
\def\ket#1{\lket #1\rket}
\def\bra#1{\lbra #1\rbra}

\def\abs#1{\lvert\lvert #1\rvert\rvert}

\def\subsubsec#1{\noindent{\undertext {#1}}}
\def\undertext#1{$\underline{\smash{\hbox{#1}}}$}

\def\calo{{\cal O}}
\def\calh{{\cal H}}
\def\calN{{\cal N}}

\def\hext{{{\cal H}_{\rm ext}}}
\def\hnear{{{\cal H}_{\rm near}}}
\def\hfar{{{\cal H}_{\rm far}}}
\def\hbh{{{\cal H}_{\rm BH}}}
\def\ha{{A}}
\def\hb{{B}}
\def\hc{{C}}

\def\dima{{|A|}}

\def\roughly#1{\mathrel{\raise.3ex\hbox{$#1$\kern-.75em\lower1ex\hbox{$\sim$}}}}
\def\ahat{{\hat a}}
\def\ahats{|\ahat\rangle}
\def\as{|a\rangle}

\def\rtt{{\sqrt2}}
\def\zhat{{\hat 0}}
\def\ohat{{\hat 1}}
\def\zhats{|\zhat\rangle}
\def\ohats{|\ohat\rangle}
\def\zs{|0\rangle}
\def\os{|1\rangle}
\def\qhat{{\hat q}}
\def\uhat{{\hat U}}
\def\vecm{{\vec m}}

\def\mthsu{\mathsurround=0pt  }
\def\leftrightarrowfill{$\mthsu \mathord\leftarrow\mkern-6mu\cleaders
  \hbox{$\mkern-2mu \mathord- \mkern-2mu$}\hfill
  \mkern-6mu\mathord\rightarrow$}
 \def\overleftrightarrow#1{\vbox{\ialign{##\crcr\leftrightarrowfill\crcr\noalign{\kern-1pt\nointerlineskip}$\hfil\displaystyle{#1}\hfil$\crcr}}}
\overfullrule=0pt

\def\comment#1{}

%
%
\lref\Pagerev{
  D.~N.~Page,
  ``Black hole information,''
[hep-th/9305040].
}
\lref\SGTrieste{S.~B.~Giddings,
  ``Quantum mechanics of black holes,''
  arXiv:hep-th/9412138\semi
    ``The black hole information paradox,''
  arXiv:hep-th/9508151.
}
\lref\Mathurrev{
  S.~D.~Mathur,
  ``The Information paradox: A pedagogical introduction,''
Class.\ Quant.\ Grav.\  {\bf 26}, 224001 (2009).
[arXiv:0909.1038 [hep-th]];  ``What the information paradox is {\it not},''  
  [arXiv:1108.0302 [hep-th]].
}
\lref\Astrorev{
  A.~Strominger,
  ``Les Houches lectures on black holes,''
  arXiv:hep-th/9501071.
}
\lref\BPS{
  T.~Banks, L.~Susskind and M.~E.~Peskin,
  ``Difficulties for the Evolution of Pure States Into Mixed States,''
Nucl.\ Phys.\ B {\bf 244}, 125 (1984)..
}
\lref\subaddref{
  H. Araki and E. H. Lieb,
  ``Entropy Inequalities,''
  Commun.\ Math.\ Phys.\  {\bf 18}, 160 (1970).
}
\lref\ADM{
  R.~L.~Arnowitt, S.~Deser and C.~W.~Misner,
 ``Canonical variables for general relativity,''
Phys.\ Rev.\  {\bf 117}, 1595 (1960).  
}
\lref\Nohiding{
  S.~L.~Braunstein and A.~K.~Pati,
  ``Quantum information cannot be completely hidden in correlations: Implications for the black-hole information paradox,''
Phys.\ Rev.\ Lett.\  {\bf 98}, 080502 (2007).
[gr-qc/0603046]. 
}
\lref\Braunetal{
  S.~L.~Braunstein, H.~-J.~Sommers and K.~Zyczkowski,
  ``Entangled black holes as ciphers of hidden information,''
[arXiv:0907.0739 [quant-ph]].
}
\lref\Braun{
  S.~L.~Braunstein,
  ``Entangled black holes as ciphers of hidden information,''
[arXiv:0907.1190 [quant-ph]].
}
\lref\PageCPT{
  D.~N.~Page,
  ``Is Black Hole Evaporation Predictable?,''
Phys.\ Rev.\ Lett.\  {\bf 44}, 301 (1980)..
}
\lref\Page{
  D.~N.~Page,
  ``Average entropy of a subsystem,''
Phys.\ Rev.\ Lett.\  {\bf 71}, 1291 (1993).
[gr-qc/9305007].
}
\lref\Banks{
T.~Banks, L.~Susskind and M.~E.~Peskin,
  ``Difficulties for the Evolution of Pure States Into Mixed States,''
Nucl.\ Phys.\ B {\bf 244}, 125 (1984)..
}
\lref\Gidrem{
  S.~B.~Giddings,
  ``Comments on information loss and remnants,''
Phys.\ Rev.\ D {\bf 49}, 4078 (1994).
[hep-th/9310101].
}
\lref\NLvC{
  S.~B.~Giddings,
  ``Nonlocality versus complementarity: A Conservative approach to the information problem,''
Class.\ Quant.\ Grav.\  {\bf 28}, 025002 (2011).
[arXiv:0911.3395 [hep-th]].
}
\lref\Gidinfprod{
  S.~B.~Giddings,
   ``Why aren't black holes infinitely produced?,''
Phys.\ Rev.\ D {\bf 51}, 6860 (1995).
[hep-th/9412159].
}
\lref\GidNel{
  S.~B.~Giddings and W.~M.~Nelson,
   ``Quantum emission from two-dimensional black holes,''
Phys.\ Rev.\ D {\bf 46}, 2486 (1992).
[hep-th/9204072].
}
\lref\SGmodels{
  S.~B.~Giddings,
   ``Models for unitary black hole disintegration,''
[arXiv:1108.2015 [hep-th]].
}
\lref\SGunit{
  S.~B.~Giddings,
  ``Black holes, quantum information, and unitary evolution,''
  Phys.\ Rev.\ D {\bf 85}, 124063 (2012).
[arXiv:1201.1037 [hep-th]].
}
\lref\SGnonlocal{
  S.~B.~Giddings,
   ``Black hole information, unitarity, and nonlocality,''
  Phys.\ Rev.\ D {\bf 74}, 106005 (2006).
  [hep-th/0605196].
}
\lref\AveryNB{
  S.~G.~Avery,
  ``Qubit Models of Black Hole Evaporation,''
[arXiv:1109.2911 [hep-th]].
}
\lref\Hawkrad{
  S.~W.~Hawking,
  ``Particle Creation By Black Holes,''
  Commun.\ Math.\ Phys.\  {\bf 43}, 199 (1975)
  [Erratum-ibid.\  {\bf 46}, 206 (1976)].
}
\lref\Hawkunc{
  S.~W.~Hawking,
  ``Breakdown Of Predictability In Gravitational Collapse,''
  Phys.\ Rev.\  D {\bf 14}, 2460 (1976).
}
\lref\HaPr{
  P.~Hayden, J.~Preskill,
  ``Black holes as mirrors: Quantum information in random subsystems,''
JHEP {\bf 0709}, 120 (2007).
[arXiv:0708.4025 [hep-th]].
}
\lref\Hayden{
  P. Hayden, R. Jozsa, D. Petz, and A. Winter,
  ``Structure of states which satisfy strong subadditivity of quantum entropy with equality''
  Commun.\ Math.\ Phys.\  {\bf 246}, 359 (2004).
[arXiv:quant-ph/0304007].
}
\lref\subadfact{
  M. A. Nielsen and I. L. Chuang,
  {\sl Quantum Computation and Quantum Information,}
   Cambridge University Press (2000).
}
\lref\LPSTU{
  D.~A.~Lowe, J.~Polchinski, L.~Susskind, L.~Thorlacius and J.~Uglum,
  ``Black hole complementarity versus locality,''
  Phys.\ Rev.\  D {\bf 52}, 6997 (1995)
  [arXiv:hep-th/9506138].
}
\lref\Susstroub{
  L.~Susskind,
  ``Trouble for remnants,''
[hep-th/9501106].
}
\lref\Subadd{
  E.~H.~Lieb and M.~B.~Ruskai, ``Proof of the strong subadditivity of quantum-mechanical entropy,"
  J.\ Math.\ Phys.\ {\bf 14}, 1938 (1973).
}
\lref\MaPo{
  I.~Heemskerk, D.~Marolf and J.~Polchinski,
  ``Bulk and Transhorizon Measurements in AdS/CFT,''
[arXiv:1201.3664 [hep-th]].
}
\lref\fuzz{
  S.~D.~Mathur,
   ``The information paradox: conflicts and resolutions,''
[arXiv:1201.2079 [hep-th]].
}
\lref\SeSu{
  Y.~Sekino and L.~Susskind,
  ``Fast Scramblers,''
JHEP {\bf 0810}, 065 (2008).
[arXiv:0808.2096 [hep-th]].
}
\lref\SusskindAP{
  L.~Susskind,
  ``Addendum to Fast Scramblers,''
[arXiv:1101.6048 [hep-th]].
}
\lref\isodef{
  P. A. Fillmore,
  ``The shift operator,''
  The American Mathematical Monthly  {\bf 81}, 7, 717 (1974).
}
\lref\LashkariYI{
  N.~Lashkari, D.~Stanford, M.~Hastings, T.~Osborne and P.~Hayden,
  ``Towards the Fast Scrambling Conjecture,''
[arXiv:1111.6580 [hep-th]].
}
\lref\kak{R. R. Tucci,
  ``An introduction to Cartan's KAK decomposition for qc programmers,''
[arXiv:0507171v1 [quant-ph]]
}
\lref\UQM{
  S.~B.~Giddings,
  ``Universal quantum mechanics,''
Phys.\ Rev.\ D {\bf 78}, 084004 (2008).
[arXiv:0711.0757 [quant-ph]].
}
\lref\Haag{
  R.~Haag,
  {\sl Local quantum physics: Fields, particles, algebras,}
Berlin, Germany: Springer (1992) 356 p. (Texts and monographs in physics).
}
\lref\holoST{
  T.~Banks,
  ``Holographic Space-Time: The Takeaway,''
[arXiv:1109.2435 [hep-th]].
}
\lref\wabhip{
  S.~B.~Giddings,
  ``Why aren't black holes infinitely produced?,''
Phys.\ Rev.\ D {\bf 51}, 6860-6869 (1995).
[hep-th/9412159].
}
\lref\fuzz{
  S.~D.~Mathur,
  ``Fuzzballs and the information paradox: A Summary and conjectures,''
[arXiv:0810.4525 [hep-th]].
}
\lref\QBHB{
  S.~B.~Giddings,
  ``Quantization in black hole backgrounds,''
Phys.\ Rev.\ D {\bf 76}, 064027 (2007).
[hep-th/0703116 [HEP-TH]].
}
\lref\BHMR{
  S.~B.~Giddings,
  ``Black holes and massive remnants,''
Phys.\ Rev.\ D {\bf 46}, 1347 (1992).
[hep-th/9203059].
}
\lref\Holog{G.~'t Hooft,
  ``Dimensional reduction in quantum gravity,''
[gr-qc/9310026]\semi
 L.~Susskind,
  ``The World as a hologram,''
J.\ Math.\ Phys.\  {\bf 36}, 6377 (1995).
[hep-th/9409089].
}
\lref\GiddingsPT{
  S.~B.~Giddings and M.~Lippert,
  ``Precursors, black holes, and a locality bound,''
Phys.\ Rev.\ D {\bf 65}, 024006 (2002).
[hep-th/0103231].
}
\lref\LQGST{
  S.~B.~Giddings,
  ``Locality in quantum gravity and string theory,''
Phys.\ Rev.\ D {\bf 74}, 106006 (2006).
[hep-th/0604072].
}

\Title{
\vbox{\hbox{NSF-KITP-12-098}}
\vbox{\baselineskip12pt  
}}
{\vbox{\centerline{Quantum information transfer and}\centerline{models for black hole mechanics}
}}

\centerline{{\ticp 
Steven B. Giddings\footnote{$^\ast$}{Email address: giddings@physics.ucsb.edu}${}^{a,b}$ and Yinbo Shi\footnote{$^\dagger$}{Email address: yshi@physics.ucsb.edu}${}^a$
} }
\centerline{\sl ${}^a$Department of Physics}
\centerline{\sl and}
\centerline{\sl ${}^b$Kavli Institute for Theoretical Physics}
\centerline{\sl University of California}
\centerline{\sl Santa Barbara, CA 93106}
\vskip.10in
\centerline{\bf Abstract}
General features of information transfer between quantum subsystems, via unitary evolution, are investigated, with applications to the problem of information transfer from a black hole to its surroundings.  A particularly direct form of quantum information transfer is ``subsystem transfer," which can be characterized by saturation of a subadditivity inequality.  We also describe more general unitary quantum information transfer, and categorize different models for black hole evolution.  Evolution that only creates paired excitations inside/outside the black hole is shown not to extract information, but information-transferring models exist both in the ``saturating" and ``non-saturating" category.  The former more closely capture thermodynamic behavior; the latter generically have enhanced energy flux, beyond that of Hawking.

\vskip.3in
\Date{}

\newsec{Introduction}

The black hole information problem\foot{For reviews, see \refs{\Pagerev\SGTrieste\Astrorev-\Mathurrev}.} appears to be a central conceptual problem in a quantum formulation of gravity.  Our existing framework for  physics - local quantum field theory (LQFT) -- has been argued to predict that black hole evaporation either leads to violation of quantum mechanics\Hawkunc\ and energy conservation\refs{\BPS} , or to black hole remnants with unboundedly large number of internal states, producing catastrophic instabilities.\foot{See, {\it e.g.}, \refs{\wabhip,\Susstroub}.}

If one assumes that quantum mechanics is valid, without unphysical instabilities, this apparently contradicts the locality property of LQFT, and thus calls for a different underlying quantum framework.  In quantum mechanics, this should be given in a Hilbert space description.  Such a framework should then reproduce LQFT as an excellent approximation in familiar circumstances, {\it e.g.} those avoiding ultra-planckian collisions.  This fits into a picture where the fundamental quantities are defined as quantum objects, such as states in Hilbert space, and not in terms of spacetime. 

Additional structure is needed to characterize the physics; a particular problem is that of recovering locality to an excellent approximation.  One way to define a basic notion of localization, in such a framework, is by specifying smaller tensor factors of a given Hilbert space.  These can be thought of as corresponding to different ``regions."  Indeed, in LQFT, the field operators localized to a given region produce such a factor structure, underlying the algebraic approach to LQFT\refs{\Haag}.\foot{Banks\refs{\holoST} has also explored using tensor factor structures to give a {\it holographic} description of space time.}    Thus, a proposal is that part of the basic framework for gravity is a network of tensor factors\SGunit.

In this approach the basic ``stuff" is quantum information, and it is conserved under unitary quantum-mechanical evolution, defined in an appropriately general sense.  In LQFT, locality also constrains such evolution, and we likewise expect constraints here.  

As a concrete example, one can explore these ideas in the case of black hole evolution.  A basic hypothesis is that we should think of the black hole and its surroundings as corresponding to subsystems of a larger quantum system, yielding a tensor factor structure.  The problem, then, is to understand unitary evolution of the combined system.  Part of this problem then becomes a more generic problem in quantum information theory:  characterizing unitary information transfer between two subsystems.

When a black hole evaporates and shrinks, one expects the number of internal states of its Hilbert space to also decrease, to avoid the problems noted above.  Thus, we consider unitary evolution where the product structure also evolves with time.   As this happens, quantum information must transfer from the black hole states to the exterior states.

We wish to further characterize and constrain the evolution relevant for black holes, and more generally quantum gravity.  In addition to basic information theory constraints, which we will explore, we also wish to describe a framework that matches onto an LQFT description in appropriate circumstances.  One might expect that in particular this means there is a good LQFT approximation for states outside the black hole, and even for states that an observer would measure falling into the black hole.  The requirement, to approximately match LQFT evolution about a semiclassical geometrical background, appears to be a powerful constraint -- indeed it is not at the outset clear it can be satisfied, in a consistent general framework.

Another way to describe the approach we take is to think of it as an {\it effective quantum theory} description of black holes and their evolution.  A black hole is expected to have a finite number of states, and we can parameterize unitary evolution describing interaction with its surroundings.  However, since there are a number of powerful constraints on this evolution, the approach is potentially more powerful than simply an effective description.  To motivate this statement, recall that the problem of parameterizing quantum evolution consistent with Poincar\'e invariance and locality essentially produces the structure of LQFT.  Likewise, here, we might expect to have sufficient constraints to learn a great deal about the more basic structure of the theory.

In outline, the next section describes more details of the  black hole states and evolution, setting up the essential problem, and also constraining possible unitary evolution.  Section three focuses on general quantum information-theoretic results characterizing information transfer between subsystems and is largely independent of the black hole story; those interested primarily in information-theoretic issues should read this section first, consulting the other sections for cultural references.  In particular, we characterize evolution in terms of a minimal form -- ``subsystem transfer," which saturates a subadditivity inequality -- and departures from that, and also compare the role of such transfer to that of scrambling.  Section four then extends these basic ideas and constraints into the black hole context, and in particular investigates existing classes of models for such evolution.  We also discuss the question of whether physical constraints imply evolution that is close to saturating subsystem transfer.  The appendices contains further technical details and proofs.

\newsec{Framework}

If nature is quantum-mechanical, at a minimum\refs{\UQM} we expect it to be described in terms of a Hilbert space of quantum states.  In LQFT, this space of states is supplied by a Fock space construction or interacting generalization.  However, it has been argued (see \refs{\BHMR\Holog\GiddingsPT\SGnonlocal-\LQGST}) that no such local description is consistent with quantum mechanics together with basic properties of gravity, and in particular black holes.  

Therefore, we seemingly need a quantum theory that doesn't originate in LQFT.  However, there are strong constraints -- one  being the statement that LQFT emerges as an excellent approximation in familiar circumstances.  A generic quantum mechanical system, even with sufficiently large Hilbert space, would not  exhibit this behavior.  A particular constraint -- though one which we expect to be subtly violated -- is that of spacetime locality.  Generic nonlocality contradicts our experience, and treated as a modification of a quantum field theory framework, leads to trouble with causality, and consequent paradoxes.  A difficult question is how to achieve {\it approximate} locality, without having the precise locality of LQFT.

Ref.~\refs{\SGunit} proposed that a more general structure, implementing a coarser notion of localization, is provided by a Hilbert space with certain tensor factors.  Specifically, the tensor factors might be thought of as associated with states in different ``regions" of spacetime.   Such a structure arises with Fock space of LQFT, but clearly can be more general.  In addition, a full statement of approximate locality involves restriction of the unitary evolution, so that ``distant" elements of the tensor factor structure don't strongly interact.  Ref.~\refs{\SGunit} proposed that these elements could provide a framework for a complete theory of quantum gravity.

If such a structure is relevant to quantum gravity, it should in particular supply a description of quantum evolution of a black hole.  Ref.~\refs{\SGmodels} gave illustrative simple models for such unitary evolution, on a restriction of the Hilbert space, and ref.~\refs{\SGunit} proposed a more general description of the possible Hilbert space structure, and unitary evolution, for describing black holes.  This paper will explore further constraints on such evolution, arising from various physical and mathematical criteria.  In order to do so, we first review aspects of the Hilbert space structure described in \refs{\SGmodels,\SGunit}.  

\subsec{Hilbert spaces and unitary evolution for black holes}

Consider the space of states of a black hole, interacting with its surroundings, in, {\it e.g.}, asymptotically flat space.  We will assume that this is a Hilbert space, with states contained in a tensor product
\eqn\BHdecomp{{\cal H}\subset {\cal H}_{BH}\otimes {\cal H}_{ext}\ ,}
corresponding to a description at a particular ``time."\foot{While we expect to have more general notions of time, for simplicity, this may be taken to be time at infinity.  Then in a geometrical description there is the question of choosing the particular time slice.  In the present framework, we expect changes of this slice could correspond to unitary equivalences, as briefly outlined in ref.~{\SGunit}.}  This is a non-trivial assumption about the quantum mechanical configurations of the system, but we deem it as plausible and worth exploring.

The description at a different time is related by a unitary operator.  More precisely, this evolution map may change the factors in \BHdecomp, and in particular their dimensions.  But, we assume that it is one-to-one on the image of physical states $\cal H$, and preserves the inner product.  These thus preserve quantum information; knowledge of the current state allows postdiction of prior events.  While technically such maps are only {\it isometries}\refs{\isodef}, we refer to them as ``unitary."

Thus, most generally we are describing interacting quantum subsystems of a larger system, such that the size of the subsystems can change through evolution.  
In a pure state, the ``missing" information from one subsystem is given by the von Neuman entropy of its density matrix, $\rho_i$
\eqn\ent{S_i = -Tr(\rho_i \ln \rho_i),}
One key feature is that $S_i\leq \ln$ dim$(\calh_i)$, which allows one to place a lower bound on various subsystems.

We expect additional structure in order to capture the physics of black holes.  First, while LQFT evolution contradicts unitarity\refs{\Hawkunc}, we do expect the evolution of low-energy states of ${\cal H}_{ext}$ far from the black hole to have an excellent LQFT description.  Moreover, for a large black hole, we expect a good approximate LQFT description of some features of the nearby  external states, and of the states ``inside" the black hole -- for example of measurements of an infalling observer, before collision with the strong curvature region.  

For unitarity's sake, we do however expect possible departures from a LQFT description for the black hole and near states and their evolution.  We will make the apparently reasonable assumption that the only significant departures affect these two subystems, and thus further divide ${\cal H}_{ext}$ into ${\cal H}_{near} \otimes {\cal H}_{far}$.  Concretely, we don't expect unitary evolution of a solar-mass black hole here to nonlocally relay information to Alpha-Centauri -- although we propose that small departures from LQFT are possible on the scale corresponding to the Schwarzschild radius, $R\sim 1\ km$, under appropriate circumstances.  Specifically, we expect significant modifications of ${\cal H}_{BH}$, and assume that the unitary evolution coupling this space with the black hole ``atmosphere" ${\cal H}_{near}$ departs from that of LQFT, but that the couplings of ${\cal H}_{near}$ with  ${\cal H}_{far}$ are for practical purposes well-approximated by LQFT.

\subsec{LQFT evolution}

If we seek a minimal departure from LQFT, let us first recall how it  describes BH evolution, and, following \refs{\SGmodels,\SGunit}, ask what modifications may be needed.   

Time evolution can be described in the ADM formalism \refs{\ADM}.  Consider perturbations about a spherically-symmetric metric, 
\eqn\admrad{ds^2 = -N^2 dT^2 + g_{xx} (dx+N^x dT)(dx+N^x dT) + r^2(T,x)d\Omega^2\ }
where $N$ and $N^x$ are the usual lapse and shift functions, respectively.  Here a choice of time-slicing has been made; $T$ labels the constant time slices and $x$ is a coordinate parameterizing the radial direction along the slice.  Different time slicings are possible; nice slices\refs{\LPSTU} clearly exhibit the tension between LQFT and unitarity.  An explicit construction of such slices is given in \SGunit, in the approximation of static geometry.  These slices asymptote to constant Schwarzschild-time slices at infinity, and asymptote to a constant radius inside the horizon, thus avoiding the singularity. 

LQFT can be set up by quantizing on this slicing.  It is simplest to consider a scalar quantum field, although other fields can be treated, including metric perturbations.  
In the scalar case, we expand the field in a basis of mode functions, and creation and annihilation operators
 \eqn\tdexpan{\phi(x,T,\Omega)=\sum_{il\vecm}\left(a_{il\vecm} u_{il}(x) {Y_{l\vecm}(\Omega)\over r^{D/2-1}} + {\rm h.c.}\right)\ .}
It is particularly helpful if the mode functions are chosen to be approximately localized in position and momentum, subject to uncertainty-principle constraints.  For example, one such construction is the windowed Fourier transform\refs{\Hawkrad,\GidNel,\SGunit}
\eqn\uki{u_{ja} = {1\over \sqrt\epsilon} \int_{j\epsilon}^{(j+1)\epsilon} dk  e^{ik(x-2\pi a/\epsilon)}\ ,\ {\tilde u}_{ja} =  {1\over \sqrt\epsilon} \int_{j\epsilon}^{(j+1)\epsilon} dk  e^{-ik(x-2\pi a/\epsilon)}\ ,}
where $\epsilon$ is a resolution parameter and $j$, $a$ index the localization in the radial momentum and radial position, respectively.  Clearly other approximately localized bases exist. 

Such localized modes give us a way to decompose the Hilbert space into $\calh(T) = \hbh(T)\otimes  {\cal H}_{near}(T) \otimes {\cal H}_{far}(T)$ at a given time.  In particular, we think of states associated with modes localized within a few times the Schwarzschild radius, but outside the horizon, as comprising ${\cal H}_{near}(T)$.  This decomposition changes with time -- as noted in our general discussion of \BHdecomp.

The Hawking radiation can be exhibited in terms of a particular entangled state in ${\cal H}_{BH}\otimes {\cal H}_{ext}$.  An important condition for determining this state is that the infalling observer sees no high-momentum excitations near the horizon -- these modes are in their vacuum.  But, evolution of this state produces correlated pairs of excitations, with one partner of each pair escaping as a quantum of Hawking radiation, and one falling into the BH interior.  Since the high-momentum modes are in their vacuum, it is useful to introduce a high-momentum cutoff to describe this state.  Specifically, focusing on the outgoing, near-horizon modes, this state takes the form\SGunit 
\eqn\sqopversc{|\psi\rangle_{HR} =\prod_{jl}\prod_a^{A(T)} \left(S |\hat 0\rangle|0\rangle\right)_{jal} |0\rangle_{A(T)}\ .}

Here $a<A(T)$ is needed for the high-momentum cutoff, with $A(T)=\epsilon(T+kR)/(2\pi)$, and $k$ a constant determined by the cutoff momentum.  The corresponding short-wavelength modes are in their vacuum, $|0\rangle_{A(T)}$.  $S$ is a squeeze operator, of the form 
\eqn\Sqedef{S_{jal} = \exp\left\{{z(\omega_j)} \left( {\hat b}_{jal}^\dagger b_{jal}^\dagger -  {\hat b}_{jal} b_{jal}\right)\right\}\ ,}
with
\eqn\zodef{\tanh z(\omega) = e^{-\beta \omega/2}\ .}
In keeping with conventions used in \refs{\GidNel,\Mathurrev}, hatted quantities correspond to inside states.
This construction is particularly explicit in two-dimensional models\GidNel.  

The Hawking radiation can be described by a density matrix, formed by  tracing out the black hole interior states.   This results in a thermal density matrix\foot{This discussion neglects reflection -- see \SGunit\ for more discussion.}
\eqn\densmat{\rho(T) = {1\over Z} \sum_{\{n_{jal}\}, a<A(T)}  e^{-\beta H} |\{n_{jal}\}\rangle\langle\{n_{jal}\}|\ }
where $n_{jal}$ are mode occupation numbers.
As $T$ grows, so does $A(T)$, and the entropy \ent\ of \densmat\ grows.  If one considers evolution to time scales comparable to the evaporation time, $T_{evap}\sim R S_{BH}$, with $S_{BH}$ the Bekenstein-Hawking entropy, the von Neumann entropy will be of size $S_{BH}$.  This represents the missing information.  Since local evolution apparently forbids its escape while the black hole is larger than the Planck scale (and leaving it behind in a remnant leads to other problems, see {\it e.g.} \refs{\wabhip, \Susstroub}), this parameterizes the unitary violation of LQFT.  

An essential aspect of the problem arises from the growth of the internal Hilbert space ${\cal H}_{BH}$ with time, and this in turn is forced by unitary local evolution because information is forbidden by locality from escaping.  

\subsec{Models for evolution}

If the dynamics can indeed be described in terms of subsystems \BHdecomp, we need unitary evolution such that the dimension of ${\cal H}_{BH}$ shrinks, while unitary evolution transfers its information to ${\cal H}_{ext}$.  Here, apparently, evolution must depart from that of LQFT.  A basic goal of this paper is to refine understanding of possible such evolution.

Important constraints  were outlined in \SGunit.  First, we seek Hilbert spaces and evolution with ``least possible" deviation from LQFT, which we expect to work well in familiar circumstances.  One reasonable expectation is that black holes have familiar general features, both for outside and infalling observers.  We might also expect that at least at the coarse-grained level, and at sufficiently early times, black holes evaporate approximately thermally as predicted by Hawking.  These, together with the demand of unitary evolution of the black hole, with shrinking ${\cal H}_{BH}$, apparently provide nontrivial constraints.

Indeed, in characterizing the evolution we also use constraints from information theory.  Information theory traditionally deals with finite dimensional spaces, but the LQFT Hilbert space (Fock space) is infinite dimensional.   Many results in information theory extend to infinite dimensional Hilbert spaces, but the proofs are often substantially more obfuscated or non existent\footnote{$^{\dag}$}{Fortunately, strong subadditivity remains true in infinite dimensions \refs{\Subadd}, as does Klein's inequality, which is used to prove strong subadditivity}. 

However, there are well-motivated reasons to expect that for our purposes we only need to consider finite-dimensional Hilbert spaces.  First, as has been noted, we seek a description where $\hbh$ is finite-dimensional.  Secondly, we have suggested an apparently reasonable assumption  that it only interacts significantly with the black hole atmosphere $\hnear$; usual LQFT evolution then carries the information outward (or, brings information in from $\hfar$).  Since $\hnear$ is the space of states corresponding to the region from the horizon to a few times larger radius, the LQFT description of this space is finite-dimensional in the presence of a UV cutoff.   In order not to introduce major deviations from the Hawking radiation -- which would be seen by an infalling observer as high-energy particles -- we might expect modifications of LQFT only to affect excitations at wavelengths longer than such a cutoff, say $A(T)$ of \sqopversc.  Thus, the unitary information transfer takes place between finite-dimensional Hilbert spaces.

In fact, an even stronger possible condition\refs{\SGmodels,\SGunit} is that the departures from LQFT only affect quanta seen by infalling observers to have energies $<K/R$, with $K$ a modest number, say $K<5$.  This makes such modifications appear very innocuous to infalling observers. In this case, the dimension of the relevant part of $\hnear$ is correspondingly small, with $\sim K^2/(2\pi)$ modes.  

Indeed, the basic features we have described suggest simplified toy models for black hole evolution, and such toy models have been explored in \refs{\Mathurrev,\SGmodels,\AveryNB,\SGunit}.  Specifically, the thermal factor with temperature $T\sim1/R$ tells us that quanta with asymptotic energies $\gg 1/R$ have exponentially suppressed amplitudes, and gray body factors suppress emission with energies $\ll1/R$.  Indeed, in practice it is useful to take the arbitrary resolution parameter in \uki\ to be $\epsilon\sim 1/R$.  Then, we find that one particle with energy $\sim 1/R$ is emitted for each time $\sim R$.  The simplest model\refs{\Mathurrev} forgets all but occupation number zero or one of one such mode, and replaces the thermal factor by one; in this case, evolution through a time $\sim R$ maps the initial state $\ket{\phi}$ of $\hbh\otimes\hext$ by
\eqn\mathmod{\ket{\phi} \rightarrow \ket{\phi} { \ket{\hat00} + \ket{\hat11}\over \sqrt 2}\ .}
This is a simplified form of evolution corresponding to shifting the cutoff in \sqopversc;  the hatted/unhatted qubits correspond to modes just inside/outside the horizon.  Such toy qubit models can be generalized, and their generalizations can be used to explore modifications to and information-theoretic constraints on evolution.

\subsec{Paired states, a no-go theorem, and Schr\"odinger's cat in a black hole}

The evolution of eq.~\mathmod\ corresponds to an increase of the entropy of the external state of one bit per time step, mirroring the more general statements made below eq.~\densmat.  We would like to understand what kinds of modifications to evolution avoid the increase in entropy, and in fact reduce the entropy of the external state.  

Note a prominent feature of the Hawking state is the pairing between internal and external quanta, seen in eqs.~\Sqedef\ and \mathmod.  Indeed, this pairing is part of an explanation for why the infalling observer sees nothing violent:  it can be shown\refs{\SGnonlocal} that while interactions between that observer and individual blue-shifted Hawking particles inside or outside the horizon can be large, there are cancellations between the interactions with the inside and outside modes.  

This suggests considering modifications that retain this pairing.  Ref.~\Mathurrev\ considers small admixtures of  $( \ket{\hat00} - \ket{\hat11})/\sqrt 2$, and argues that small corrections of this form to the toy Hawking evaporation \mathmod\ do not decrease the entropy of the external state.

In fact, there is a much more general result, that does not rely on smallness of corrections, but only on this pairing property.  Specifically, begin with a state of the form \sqopversc,
which is a linear combination of a countable number of basis states.  Because they are countable, they can then be well ordered in some manner.  An arbitrary black hole pure state  (including matter that made the original black hole, infalling Hawking particles, and outgoing Hawking particles) can be written as 
\eqn\a{ \ket{\phi} = \sum_{i,j}C'_{i,j}\hat\psi'_i\chi'_j }
where $\hat\psi'_i$ and $\chi'_j$ are orthonormal bases for $\hbh$ and $\calh_{ext}$, respectively.  By choice of new bases ${\hat \psi}_i$ and $\chi_j$ for $\hbh$ and $\calh_{ext}$, respectively, this can be put in the Schmidt decomposed/singular value form
\eqn\schmidt{
\ket{\phi} = \sum_{k}C_k\hat\psi_k\chi_k} 
where for each $k, C_k\geq0$  .  

Consider a general time evolution, in which new particles are emitted in states with internal/external pairing, $\ket{\hat nn}$.  Here the integer $n$ can either label different modes, or their occupation numbers, or even more general paired quantum numbers.
A general evolution to such states is
\eqn\asdf{\eqalign{
	& \chi_i \rightarrow \chi_i \\
	& \hat\psi_i \rightarrow \hat\psi^0_i\ket{\hat00} + \hat\psi^1_i\ket{\hat11} + \hat\psi^2_i\ket{\hat22} + ...
}}
We could also consider unitary evolution of 
the $\chi_i$.   But that does not change the analysis since we can always choose to use the evolved $\chi_i$ in the analysis (as long as this evolution is largely independent of the black hole, as we expect for  Hawking particles emitted some time ago, which have long since left the black hole vicinity).  The $\hat\psi^n_i$ are just some (generally not normalized) linear combination of $\hat\psi_i$.   Unitarity preserves norms, so for each $i$,
\eqn\d{ \sum_n\abs{\hat\psi^n_i}^2 = 1\ . }
Combining \schmidt\ and \asdf, 
the new state is
\eqn\state{ \eqalign{
	\ket{\phi}' & = \sum_{i}C_i\left(\hat\psi^0_i\ket{\hat00} + \hat\psi^1_i\ket{\hat11} + \hat\psi^2_i\ket{\hat22} + ...\right)\chi_i \\
	& = \left(\sum_{i}C_i\hat\psi^0_i\chi_i\right)\ket{\hat00} + \left(\sum_{i}C_i\hat\psi^1_i\chi_i\right)\ket{\hat11} + ...\\
	& = \Lambda^0\ket{\hat00} + \Lambda^1\ket{\hat11} + ... 
}}
with $\Lambda^n = \sum_{i}C_i\hat\psi^n_i\chi_i$.

In the case of the Hawking state, $\Lambda^n = (e^{-\beta E_n})\Lambda^0$.  The Hawking pair factors out, and  it is clear that every pair increases the entanglement entropy between the inside and outside.

Since Hawking's result is not exact, many have speculated that small corrections, contributing to many Hawking pairs, may allow information escape.  As noted, ref.~\Mathurrev\ explored this in such paired models by adding a small admixture of   $( \ket{\hat00} - \ket{\hat11})/\sqrt 2$ that could depend on the internal state of the black hole.  It then showed that the entropy increases by at least $\ln(2) -2\epsilon$ for each pair (where $\epsilon \ll 1$ is a parameter that defines the size of the perturbation), demonstrating that such small perturbations cannot restore unitarity.

The broader result states that for  {\it general} evolution of the form \state, the entropy of the external state {\it cannot decrease} -- independent of the question of smallness of the corrections.  The proof appears in appendix A.  So, one finds that the real issue is not smallness of the corrections, but the evolution into paired states of internal and  external particles.

At first glance, this may seem like an odd result.  For example, consider a unitary operator that maps
\eqn\attempt{\eqalign{ \ket{\zhat 0} &\rightarrow \ket{\zhat 0}\ ;\cr \ket{\ohat 0} &\rightarrow \ket{\ohat 1}\ . }}
A CNOT gate does this, and is known to be unitary.  It certainly seems that the outside observer can uniquely determine the initial state of the interior based on what is observed coming out.  Unfortunately, this is an illusion that stems largely from the fact that this operator is a legitimate {\it classical} cloner.  The No Cloning Theorem, of course, prohibits {\it quantum} cloning.  To see that this evolution doesn't extract the information, consider its action on the following orthogonal states: ${1\over\rtt} \left( \ket{\zhat 0} + \ket{\ohat 0} \right)$ and ${1\over\rtt} \left( \ket{\zhat 0} - \ket{\ohat 0} \right)$.  In both cases, the outside observer measures a density matrix proportional to the identity - what comes out is in fact indistinguishable from uniform noise.

These observations extend that of the earlier proven No Hiding Theorem \refs{\Nohiding}.  This states that if all density states $\rho_I$ on some subspace $I$ unitarily map to the same density state $\rho_O$ on $O$, then all the information about $I$ resides in $\calh/O$.  More intuitively, this theorem prevents generic quantum information from hiding purely in the correlations between two subsystems of a Hilbert space ({\it i.e.} local measurements in either or both Hilbert spaces reveal nothing about the information hidden).  What we've proven is stronger: even if $\rho_O$ is allowed to depend on $\rho_I$, 
$\rho_O$ does not contain the information if the evolution involves paired states.

Note parenthetically that the preceding comments connect to recent discussion of the question of measuring Schr\"odinger's cat inside a black hole\refs{\MaPo}.  If $\ket{\hat 0}$ and $\ket{\hat 1}$ represent ``live" and ``dead," respectively, then evolution \attempt\ would allow an external observer to measure whether the cat is alive or dead.  But, such evolution is not sufficient to transfer the quantum information of the state from inside  the black hole to the outside, and arbitrary measurements of the state of the cat can't be performed from measuring the outside bits.  This example thus illustrates the importance of complete quantum information {\it transfer} for unitary black hole evolution.

This discussion should make it clear that there are important constraints to be satisfied in order to restore unitarity to black hole decay, and in particular that one needs to go beyond even large departures from the Hawking result which involve paired quanta.  Classes of models that do so were given and illustrated in \refs{\SGmodels,\SGunit}, and will be discussed below.  But, a first question is how to generally characterize the type of information transfer needed, and   refine our understanding of physical constraints on unitary evolution of black holes.  We next turn to these general information-theoretic considerations.

\newsec{Characterizing Information Transfer}

Motivated by the preceding discussion, we are interested in a general characterization of quantum information transfer between subsystems of a quantum system, via unitary evolution.  In order to discuss this in a general setting, in this section we will use $\ha$ in place of the black hole Hilbert space $\hbh$, and $\hb$ in place of the external Hilbert space $\hext$ (or $\hnear$).  Thus, we consider unitary maps 
\eqn\genunit{ U: \ha\otimes\hb \rightarrow \ha'\otimes\hb' }
that transfer quantum information from a subsystem A to a subsystem B.  The information capacity of each system is given in terms of its dimension as $\ln|A|$, $\ln|B|$.  In general, we allow the dimensions of $\ha$ and $\hb$ to change.  In fact, the terminology ``unitary" is a minor abuse, as the map $U$ may not be onto $\ha'\otimes\hb' $; more precisely we consider maps that are isometries.

Making contact with standard notions of quantum information theory, each such unitary \genunit\ can be characterized as a set of quantum channels from $A \rightarrow B'$.  To see this, start with an initial density matrix $\rho = \rho_A \otimes \ket{\phi_B}\bra{\phi_B}$, with $\rho_A$ on $A$ and $\ket{\phi_B}$ a basis state on $B$, that maps to $\rho'$ under unitary action.  Each $\phi_B$ then labels a channel $Tr_B(\rho) = \rho_A \rightarrow Tr_{A'}(\rho')$.  These channels are time dependent; part of the time evolution derives from the change due to the unitary action, and part from allowing the dimensions of $A$ and $B$ to change.  

The space of unitary transformations is large, but a number  of them don't transfer information.  For example, unitaries of the form   $U_A\otimes U_B$, which can be described as {\it local unitaries}, do not do so.  Of course, one of the reasons the von Neumann entropy \ent\ is useful in characterizing information content is its invariance under such local unitaries.\foot{For a simple example, consider the transfer of information from one qubit to another.  A generic unitary acting on this system is described by SU(4).  Of its 15 generators, only 3 are nonlocal \refs{\kak}.  Of those, only 1 or 2 actually characterize information transfer.  This is a substantial simplification of the original problem of characterizing all possible unitaries.  
}

There is also a particularly simple class of transformations that do transfer information between subsystems.  For example suppose that $A$ is a tensor product, $A=A_1\otimes A_2$, with bases $\ket{i_1}$, $\ket{i_2}$ for the two factors, and let $\ket{\phi}_B$ be an arbitrary state of $B$.  Then, consider a unitary $U$ that maps to $A'=A_1$, $B'= A_2\otimes B$ via
\eqn\subspacexfer{\ket{i_1 i_2}_A \ket{\phi_B} \rightarrow \ket{i_1}_{A'} \ket{i_2\phi}_{B'}\ .}
In other words, it simply transfers the subsystem $A_2$ between the subsystems.  We will refer to such a transformation as {\it subsystem transfer}; a special case is qubit transfer.
Clearly subsystem transfer can be generalized to also include the action of local unitaries before or after the transfer.

Obviously there are other, more complicated, forms of information transfer.  We would like to better understand the features of and constraints on such transfer.

\subsec{Tracking information transfer with a reference Hilbert space}

While entropy is a useful characteristic of information transfer, more refinement is possible. Suppose there is a subsystem, say $\ha$, whose information we want to track.  To do so,  as described in \HaPr, introduce an auxiliary subsystem $\hc$ with the same dimension $\dima$  as the subsystem.  Then, choose an orthonormal basis for each and consider a maximally entangled state of $\ha$ and $\hc$:
\eqn\tracker{ \ket{\psi} = {1\over \sqrt{\dima}} \sum_{i=1}^\dima   \ket{i_A} \ket{i_C}\ .}
Evolution acting on $\ha$ is extended by trivial evolution on $C$; if $U$ is a unitary acting on $A$ (possibly together with other subsystem of the full system), 
\eqn\trackevol{U\rightarrow U \otimes I_C\ .}
The correlations with the $\ket{i_C}$ can be used to track where the quantum information in $\ha$ goes, under evolution; picturesquely, we can think of these correlations as `ropes' between these states and the corresponding states in $\ha$, or their images.   If, after evolution, there are correlations between $\hc$ and some other subsystem $\hb$ of the full Hilbert space, those characterize the quantum information transfer to that subsystem from $\ha$.

For example, suppose that we start with uncorrelated state of two subsystems $A$ and $B$; the general such state is of the form $\ket{\psi_A} \ket{\phi_B}$, and can be formed as a superposition of $\ket{i_A} \ket{\phi_B}$.  Information can be encoded in $A$ by taking different superpositions of these, and it can be transferred to  $B$ by action of a general unitary \genunit.  Thus, consider introducing the tracking state \tracker: 
\eqn\compstate{\ket{\psi} \ket{\phi_B}\in \calh_A\otimes\calh_B\otimes\calh_C\ ,}
and its corresponding density matrix $\rho_{ABC}=\ket{\psi} \ket{\phi_B}\bra{\phi_B}\bra{\psi}$.  From this, we can find the density matrices of the different subsystems, {\it e.g.} $\rho_A= \Tr_{BC}\, \rho_{ABC}$, $\rho_B= \Tr_{AC}\, \rho_{ABC}$, $\rho_{AB}= \Tr_{C}\, \rho_{ABC}$, {\it etc.}  Due to lack of correlations between $A$ and $B$, $\rho_B$ is a pure state; its entropy \ent, vanishes:  $S_B=0$.  Likewise $\rho_{AC}$ is pure, but $\rho_{AB}$ and $\rho_A$ are mixed, with entropy $S_{AB}=S_A = \ln \dima$, representing the correlations with the auxiliary subsystem $C$.

Now, evolve via a unitary \genunit, \trackevol.   If $\rho_B$ remains pure, information has not been transferred $B$.  But, if after evolution $S_B\neq 0$, correlations have been transferred to or formed with $B$.  Note that $S_{AB}$ stays fixed at $\ln \dima_0$, by unitarity of $U\otimes 1_C$.  No information transfer takes place between $A\otimes B$ and $C$: the latter is just a tool used in tracking.

While $S_B\neq0$ indicates that correlations have been formed with $B$, that does not mean information has been transferred ``out of" $A$; it could for example reside in non trivial correlated states of the two subsystems.  One way to diagnose this is to look at $S_A$.  Its decrease, representing a decrease of correlation between $A$ and $C$, is an indication of information transfer out of $A$.  Indeed, we see that $S_A$ defined in this fashion is a good measure of the amount of information in subsystem $A$.  In particular, evolution to $S_A=0$ corresponds to complete transfer of the  information from subsystem $A$ to subsystem $B$.

These entropies obey a triangle inequality\refs{\subaddref}:
\eqn\triang{|S_A-S_B|\leq S_{AB} \leq S_A+S_B\ ,}
and the rightmost inequality is the {\it subadditivity} inequality.  We can rewrite this as $S_B\geq S_{AB} - S_A$, and interpret it as saying that if correlations with $A$ are decreased by \trackevol, there is a lower bound to the increase of the correlations with $B$.  Exceeding this lower bound is caused by entanglement between $A$
and $B$.  Correspondingly, one defines the {\it mutual information} of $A$ and $B$,
\eqn\mutent{
I(A:B) = S_A + S_B - S_{AB}\ ,}
which parameterizes the correlations between $A$ and $B$.

We might ask if there is a ``minimal" form of information transfer, that produces final states saturating  the subadditivity inequality, that is, so that the mutual information $I(A:B)$ stays fixed at zero.  It turns out that there is -- and this is subsystem transfer.

\subsec{Saturation of subadditivity implies subsystem transfer}

The preceding statement takes the form of a theorem. 

{\bf THEOREM} Consider  evolution \genunit, \trackevol\ of  $\ket{\psi} \ket{\phi_B}$, where $\ket{\psi}$ is the tracker state \tracker.  Suppose that $\rho_{AB}$ after evolution  saturates subadditivity.  $U$ can then be expressed, up to local unitaries, in the canonical form \subspacexfer\ for subsystem transfer.

Saturation of subadditivity, $S_{AB}= S_A + S_B$ holds if and only if\refs{\subadfact} $\rho_{AB} = \rho_A \otimes \rho_B$.  If the eigenvalues of $\rho_A$ are $\{\rho_i\}$, and of $\rho_B$ are $\{\sigma_j\}$, then the eigenvalues of $\rho_A \otimes \rho_B$ are $\{\rho_i \sigma_j\}$.  

On the other hand, the evolution of $\ket{\psi} \ket{\phi_B}$ takes the form
\eqn\psifin{ \eqalign{
	U(\ket{\psi} \ket{\phi_B})& = {1\over\sqrt{\dima}} \left( \ket{\psi_1}\ket{1_C} + \dots + \ket{\psi_{\dima}}\ket{\dima_C} \right) \\
	& \text{with} \\
	\ket{\psi_i} & = U(\ket{i_A}\ket{\phi_B})\ ,
}}
giving the density matrix
\eqn\rhoab{ \rho_{AB} = Tr_C(U\ket{ \psi}\ket{\phi_B}\bra{\phi_B}\bra{ \psi}U^\dagger) = {1\over \dima} \left( \ket{ \psi_1}\bra{ \psi_1} + \dots \ket{ \psi_\dima}\bra{ \psi_\dima} \right) }
So, the eigenvalues of $\rho_{AB}$ are $\dima$ copies of $1/\dima$.  

This means all the nonzero eigenvalues of $\rho_A$ are the same, and the same applies to $\rho_B$.  Since we know their respective entropies, their eigenvalues must be $\dima/k$ copies of $k/\dima$ and $k$ copies of $1/k$, respectively, with $k$ an integer that divides $\dima$.

Indeed, this follows from a corollary:

{\bf Corollary:} In this context, saturation of subadditivity is equivalent to $S_B = \ln{k}$ and $S_A = \ln{\frac{\dima}{k}}$.

In one direction, this follows because $\rho_A$ and $\rho_B$ are proportional to the identity, and their respective entropies must be $\ln$ of corresponding integer dimensions.  Saturation implies that the product of these integers is $\dima$.  In the other direction, $S_A + S_B = \ln{\frac{\dima}{k}} + \ln{k} = \ln{\dima} = S_{AB}$, so subadditivity is saturated.

This then implies that $\rho_A$ is spanned by the kets/bras of an $\dima/k$ dimensional subspace of $A$, $\{\ket{\hat{1}} \dots \ket{\widehat{\dima/k}}\}$.  Similarly, $\rho_B$ is spanned by the kets/bras of a $k$ dimensional subspace of $B$, $\{\ket{1} \dots \ket{k}\}$.  Since their tensor product spans $\rho_{AB}$, $U \left( A \otimes \ket{\phi_B} \right) = \{\ket{\hat{1}} \dots \ket{\widehat{\dima/k}}\} \otimes \{\ket{1} \dots \ket{k}\}$.

Now that we have a basis for the image of $A \otimes \ket{\phi_B}$, we can apply the inverse operator $U^{-1}$ acting on the image to find a basis for $A \otimes \ket{\phi_B}$.  This new basis will in general not correspond to the original basis for $A$ mentioned in the setup.  Appropriately labeling this basis then expresses $U$ in canonical form.  To put this more concretely,
\eqn\preimage{ \matrix{
\ket{\widehat{1\otimes1}} \otimes \ket{\phi_B} = U^{-1} (\ket{\hat{1}} \otimes \ket{1} )&
{\buildrel U^{-1}\over \longleftarrow} &
\ket{\hat{1}} \otimes \ket{1} \\
\ket{\widehat{1\otimes2}} \otimes \ket{\phi_B} = U^{-1} (\ket{\hat{1}} \otimes \ket{2}) & &
\ket{\hat{1}} \otimes \ket{2} \\
& \vdots & \\
\ket{\widehat{1\otimes k}} \otimes \ket{\phi_B} = U^{-1} (\ket{\hat{1}} \otimes \ket{k} )& &
\ket{\hat{1}} \otimes \ket{k} \\
\ket{\widehat{2\otimes1}} \otimes \ket{\phi_B} = U^{-1} (\ket{\hat{2}} \otimes \ket{1}) & &
\ket{\hat{2}} \otimes \ket{1} \\
& \vdots & \\
\ket{\widehat{{\frac{\dima}{k}} \otimes k}} \otimes \ket{\phi_B} = U^{-1}( \ket{\hat{\frac{\dima}{k}}} \otimes \ket{k}) & &
\ket{\hat{\frac{\dima}{k}}} \otimes \ket{k} \\
}}
With this labeling of the new basis, $U$ is manifestly subsystem transfer:
\eqn\canonsubtrans{ \matrix{
\ket{\widehat{1\otimes1}} \otimes \ket{\phi_B} &
\rightarrow U \rightarrow &
\ket{\hat{1}} \otimes \ket{1} \\
\ket{\widehat{1\otimes2}} \otimes \ket{\phi_B} & &
\ket{\hat{1}} \otimes \ket{2} \\
& \vdots & \\
\ket{\widehat{1\otimes k}} \otimes \ket{\phi_B} & &
\ket{\hat{1}} \otimes \ket{k} \\
\ket{\widehat{2\otimes1}} \otimes \ket{\phi_B} & &
\ket{\hat{2}} \otimes \ket{1} \\
& \vdots & \\
\ket{\widehat{{\frac{\dima}{k}}\otimes k}} \otimes \ket{\phi_B} & &
\ket{\hat{\frac{\dima}{k}}} \otimes \ket{k} \\
}}
Specifically, a subsystem of dimension $k$ leaves subsystem $A$ and enters $B$. 

A basis for $A$ naturally given by the physics of the problem may not be the same as that in which the subsystem transfer takes this canonical form.  For that reason, it is nice to have a basis-independent test of whether such a basis exists, in the form of saturation of the subadditivity inequality.

We should also note what the theorem {\it does not} say.  In particular, we have kept the initial state of $B$, $\ket{\phi_B}$ fixed, though arbitrary.  This means that we have only investigated a single quantum channel, as described above.  To test a different channel, we could check  whether subadditivity is saturated for $\ket{\psi} \ket{\phi'_B}$. 
 If it is, then the map $U$ is subsystem transfer in both cases.  But, the subsystem that is transferred could be a different subsystem depending on $\ket{\phi_B}$ vs. $\ket{\phi'_B}$.  So, each channel should be checked inividually.   Furthermore, since the transferred subsystems can in general differ, action on $\ket{\psi}(\ket{\phi_B} + \ket{\phi_B'})/{\sqrt{2}}$ will not correspond to subsystem transfer. Nonetheless, this can be a useful result.

\subsec{Saturating vs.~non-saturating transfer}

While subsystem transfer is the simplest form of unitary information transfer, and as we have shown follows from saturation of the subadditivity inequality in \triang, clearly there are more general forms of information transfer that produce states not saturating this inequality.  One question is whether we expect unitary black hole evolution to be simple saturating subsystem transfer, or not.  A second question is to better understand the more general forms of evolution.  We turn first to the latter.

 \Ifig{\Fig\Tensfact}{An illustration of basic bounds on information transfer.  We assume that $S_A$ decreases linearly to zero.  $S_B$ is bounded below by the lower solid (blue) line, corresponding to $I(A,B)=0$ (saturation), and bounded above by the upper solid (red) line, corresponding to maximal nonsaturation.}{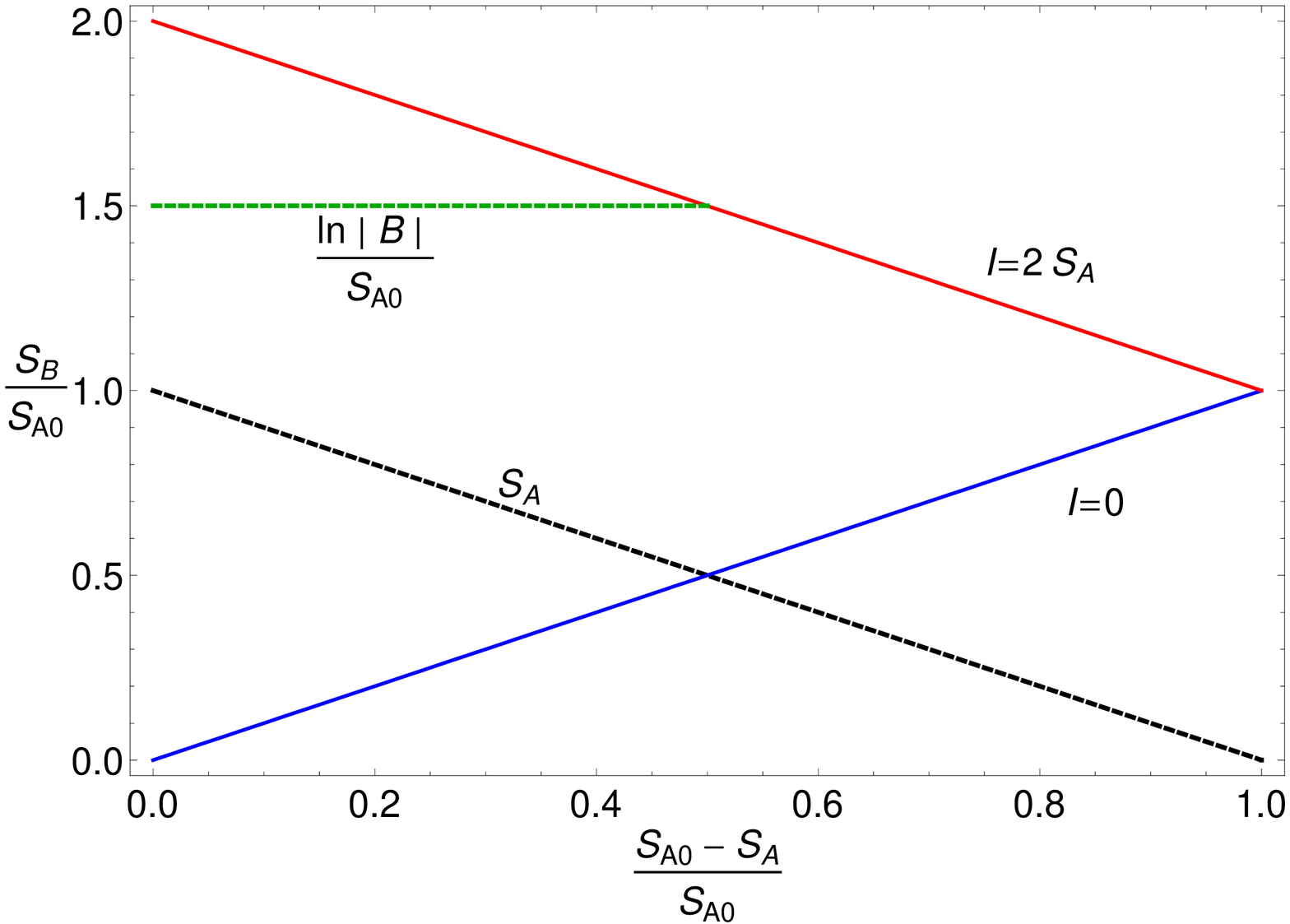}{6}

First, note that the discrete nature of subsystem transfer means that continuous evolution accomplishing it will, at intermediate stages, not saturate subadditivity.  A simple illustration of this is the continuous transfer of one bit:
\eqn\onebitc{\eqalign{\ket{\zhat 0} &\rightarrow \ket{\zhat 0}\ ;\cr
\ket{\ohat 0} &\rightarrow \cos\tau\ket{\ohat 0} + \sin\tau\ket{\zhat 1}\ .}}
At $\tau=\pi/2$, subsystem transfer has completed, but at intermediate stages the two systems are entangled in a more complicated way and $I(A:B)\neq0$.

As another illustrative example of non-saturating tranfer, consider \attempt, which we can characterize with our method of tracking information.  Here, $S_B$ increases, indicating information transfer to $B$.  But, $S_A$ does not decrease commensurately -- the information has not been transferred out of $A$.  Instead, it resides in correlations of the two systems.  

Indeed, in \attempt\  the evolution produces ``extra" excitation, in that two bits are in the excited state ``$1$."  This is another sense in which the information transfer is non-minimal.  Saturation is a condition for minimal, direct transfer.   Non-saturation corresponds to production of ``extra" entanglement, for a given amount of transferred information.

\subsubsec{Maximal nonsaturation}  

To further illustrate these considerations, we might ask whether there is a maximal departure from saturation, that is, one maximizing the mutual information  $I(A:B)$.  To begin with, a bound on this can be found as follows.  Since the combined state $\rho_{ABC}$ we consider is pure, the leftmost inequality \triang\ implies $S_B = S_{AC}$.  Then, the rightmost subadditivity inequality \triang\ implies
\eqn\subaddtwo{ S_B \leq S_A + S_C = S_A + \ln{|A|}\ . }
Thus the mutual information satisfies the bound
\eqn\maxbd{I(A:B)= S_A + S_B - S_{AB} \leq S_A + (S_A + \ln{|A|}) - \ln{|A|} = 2S_A\ .}

A unitary maximizes $I(A:B)$ iff \subaddtwo\ is saturated.  The complete state is pure, so saturation of \subaddtwo\ implies saturation of strong subadditivity\refs{\Subadd} (using $S_{AB}=S_C$, $S_{BC}=S_A$),
\eqn\ssa{S_{AB} + S_{BC} - S_{ABC} - S_B \geq 0\ .}
A lemma given in appendix B then implies that the unitary takes the simple form
\eqn\maxsuper{ \frac{1}{\sqrt{|A|}} \sum_i \ket{i_A}\ket{i_C}\ket{\phi_B} \rightarrow \ket{\psi_{AL}} \otimes \frac{1}{\sqrt{|A|}} \sum_i \ket{i_R}\ket{i_C}\ . }
Here $\hb$ must decompose as $\hb = \calh_{L} \otimes \calh_{R}$.  The state $\ket{\psi_{AL}}$ is in $\ha\otimes \calh_{L}$, and has no entanglement with $C$, and $\ket{i_R}\in \calh_{R}$.  Thus all the information has transferred out of the subsystem $A$, but entanglement between $A$ and $B$ remains.  Removing the reference subsystem, this evolution is
\eqn\maxsupertwo{ \ket{i_A}\ket{\phi_B} \rightarrow \ket{\psi_{AL}} \otimes \ket{i_R} }

In the limit that $S_A \rightarrow 0$, $\calh_{BL}$ is trivial ;
the result is saturating transfer that transfers everything.  This is consistent since $I(A:B) \leq 2S_A = 0$.  This is illustrated in fig.~1.  Of course, $S_B\leq \ln |B|$ (see green line, fig.~1), so unitary evolution is only possible if the final dimension of $B$ is as large as the initial dimension of $A$. 

Note also that deviation from saturation is bounded by the entropy $S_B$.  Specifically, for the evolution \trackevol, 
\eqn\satbound{I(A:B)= S_B + (S_A-\log|A|) \leq S_B\ .} 
 %

\subsec{Scrambling  vs. transfer}

We close this section by touching on another aspect of unitary evolution of coupled subsystems.  First, in considering general evolution \genunit, distinct forms of evolution are scrambling, and information transfer; moreover in time-dependent evolution these can have different time scales.  Scrambling of $A$\refs{\SeSu\SusskindAP-\LashkariYI} corresponds to mixing of the degrees of freedom of $A$, and thus is represented by a local unitary.  Note that its definition is basis dependent (as is the definition of a scrambling time), since it can be undone by a unitary change of basis.  Transfer of information between the subsystems $A$ and $B$, as we have described, may take place on an independent time scale.  (Of course, information transfer contributes to scrambling of the composite system.)

Both can be relevant if we want to see how fast a given degree of freedom is transferred, since that depends both on how fast it scrambles with the rest of $A$, and on how fast the transfer from $A$ to $B$ takes place.  The former dependence is because scrambling can move a given bit into a subspace that then undergoes transfer.  These basic points arise in the context of models for black hole evolution.

\newsec{Characterizing Unitary Black Hole Evolution}

We now turn to discussion of how the preceding considerations apply in the context of describing possible black hole evolution.  Let us first summarize some expectations and assumptions, following\SGunit.

First, we assume unitary evolution of the form \genunit, coupling subsystems corresponding to the black hole and its environment:
\eqn\genunitbh{ U: \hbh\otimes\hext \rightarrow \hbh'\otimes\hext'\ . }
We expect a sequence of such transformations, which might for example be parameterized by a quantity identified as ``time at infinity."

We assume that $\hbh$ decreases in dimension with evolution.  One natural proposal is that $\ln|\hbh|$ is equal to the Bekenstein-Hawking entropy $S_{BH}(M)$ corresponding to the decreasing mass of the black hole, although one may wish to consider more general time dependence.  This requires a significant departure from the semiclassical picture, since the latter describes many more states of the black hole.  This can be seen by starting with a black hole of mass $M_0$, and describing it in a nice-slicing\refs{\LPSTU,\SGunit}.  After evaporation to $M\ll M_0$, in the nice slice description one has $\calo(\exp\{S(M_0)\})$ internal states, correlated with the outgoing Hawking radiation.  

Another apparently reasonable assumption is that the external Hilbert space lies in a decomposition 
\eqn\extdecome{\hext \subset \hnear\otimes\hfar\ .}
The idea behind this is that the states and evolution on $\hfar$ are described by LQFT, as long as we consider low energy states without strong gravity effects.  Evolution of the ``black hole atmosphere" $\hnear$ may depart from that of LQFT, in particular through couplings to the states of $\hbh$.  A simplest alternative to consider is that the states of $\hnear$ are otherwise well-approximated by LQFT, although other alternatives might be considered, for example in proposals with large departure from semiclassical black hole geometry near the horizon\refs{\BHMR,\Holog,\fuzz}.  Likewise, a simplest alternative is that the couplings between $\hnear$ and $\hfar$ are well-approximated by LQFT evolution.

A key question is the evolution of $\hbh$, and its coupling to $\hnear$.  In LQFT, this evolution does not allow quantum information transfer from $\hbh$ to $\hnear$, and this results in buildup of states in $\hbh$.  Thus, LQFT evolution needs to be modified, along with the description of $\hnear$, noted above.  Nonetheless, one might seek a ``most conservative," minimal departure from LQFT in describing this evolution.  For example, we might expect the states of an infalling observer and her immediate surroundings to be well-described by LQFT, until either they impact strong curvature, in some gauge choices, or until a long time has elapsed in other gauges such as the nice slicings.  But, ultimately, unitary decrease in the size of $\hbh$ requires information transfer to $\hext$, and this is apparently outside a LQFT description.  While such evolution seems to violate locality, it does not necessarily violate causality\refs{\NLvC}.\foot{A brief explanation of this is that while in Minkowski space, Lorentz symmetry transformations can convert evolution outside the light cone into evolution backwards in time, the global symmetries of  a black hole background do not include such transformations -- the black hole can be thought of as choosing a frame.}

Thus, unless motivated otherwise by other compelling considerations, we seek unitary evolution with minimal deviation  from LQFT.  Basic aspects of the semiclassical approximation are the presence of the horizon, and that the atmosphere is essentially featureless to an infalling observer; departure from this would seem surprising.  There is potential tension between this statement and the statement that information is transferred into $\hnear$; for example, transfer into highly blueshifted Hawking modes would lead to a large departure from the Hawking state, and potentially painful effects for infalling observers.  But, in the context of more general unitary evolution, we can examine the proposal\refs{\SGmodels,\SGunit} that  information transfer only occurs to ``soft" states of $\hnear$, that is, those that correspond to quanta of moderate wavelength, and thus to particles that an infalling observer doesn't see as highly energetic.  Then, with the required information transfer rates, the alteration of the Hawking state can have minimal impact on these observers.

One expects that other physical requirements should be added to this list (see {\it e.g.} \SGunit), but we next turn to discussion of some simple models of unitary black hole evolution exhibiting some of these features, and the considerations of the preceding discussion.

\subsec{Page's random unitaries, and subadditivity}

An early description of one kind of unitary evolution is Page's \refs{\Page}.  This analysis assumes that there are black hole and radiation subsystems, with respective dimensions  $|\hbh|=\exp\{S_{BH}\}$ and $|{\cal H}_{\rm rad}|=\exp\{S_{rad}\}$, and that these dimensions change so that 
\eqn\constdim{N=|\hbh|\times|{\cal H}_{\rm rad}| ,}
 remains constant.
Page does not describe more details of the states or dynamics, but does consider properties of a random pure state in the product Hilbert space, resulting from random unitary evolution.  A particular question is the entanglement entropy of such a state, as a function of the changing dimension $|\hbh|$.  Under these conditions, he finds that the entropy of the radiation subsystem increases, until the dimensions of the two subsystems become comparable, after which the entropy of the radiation subsystem decreases to zero.

Since the dimension of the full Hilbert space remains constant under the unitary evolution, and initially the radiation system is empty, we see that what is being assumed is an example of subsystem transfer:  degrees of freedom (or subsystems) are being directly transferred from the BH subsystem to the radiation subsystem.  In particular, the entropies are at the maximum possible, given by the dimensions of the subsystems, if all states are tracked with the auxiliary subsystem.

\subsec{Unitary models approximating LQFT}

One would like to go further, and give a more detailed description of the internal and external Hilbert spaces and their evolution, that ultimately fits in a consistent framework for quantum gravity, and matches LQFT evolution in appropriate circumstances.  Specifically, we might investigate how these could more-or-less closely match semiclassical expectations, such as benign evolution for infalling observers, and radiation that approximates Hawking's.  

In the context of qubit models, \SGmodels\ provides such examples, and \SGunit\ explains how these are generalized to more realistic degrees of freedom.  

One type of evolution is described in (4.18) of \SGunit, and generalizations.  The simplified qubit version of this kind of evolution takes the form
\eqn\evoltwo{\eqalign{\zhats\zhats \ahats \as &\rightarrow \uhat \ahats\otimes \calN\left(\zhats \zs+e^{-\beta\omega/2} \ohats\os\right)\otimes U \as\ ,\cr  \zhats\ohats \ahats \as &\rightarrow \uhat \ahats\otimes\zhats\os\otimes U\as\ ,\cr \ohats \zhats \ahats \as &\rightarrow \uhat \ahats\otimes\ohats\zs\otimes U\as\ ,\cr \ohats\ohats \ahats \as &\rightarrow \uhat \ahats\otimes \calN\left(e^{-\beta\omega/2}\zhats \zs- \ohats\os\right)\otimes U \as\  }} 
 for a single time step transferring one bit of information, with normalization factor
\eqn\ndef{ \calN=(1+e^{-\beta\omega})^{-1/2}\ .}

This evolution saturates subadditivity.  This can be seen directly by defining $\ket{\tilde{0}}\ket{\tilde{0}} = {\cal N}(e^{-\beta\omega/2}\ket{\ohat}\ket{\ohat} + \ket{\zhat}\ket{\zhat})$ and $\ket{\tilde{1}}\ket{\tilde{1}} = \calN( e^{-\beta\omega/2}\ket{\zhat}\ket{\zhat}- \ket{\ohat}\ket{\ohat})$.  This basis exhibits the evolution of \evoltwo\ as subsystem transfer of one qubit.
This can also be seen  more indirectly by noticing that this map includes all possible states and preserves dimension.  One way to describe this model is to say that the usual Hawking pair that appears arises from an initial ``vacuum" state of the black hole.  But, as the interior piles up with other states, either partners of previously-emitted Hawking particles, or from infalling matter, the black hole behavior changes.  In the absence of rapid scrambling (which can be described via ${\hat U}$), this model will take quite some time before the internal space starts coming out, prolonging semiclassical behavior.  With rapid scrambling, the information begins to come out on the scrambling time scale, and in this sense the semiclassical approximation breaks down equally quickly.  This discussion illustrates the separation between the roles of information transfer, and scrambling.

A second type of model is (4.19) of \SGunit, and generalizations, whose simplified qubit form is
\eqn\evolthree{ |\qhat_1\qhat_2\rangle \ahats \as \rightarrow \uhat \ahats\otimes \calN\left(\zhats \zs+ e^{-\beta \omega/2} \ohats\os\right)\otimes|\zhat'\zhat''\rangle|q_1'q_2''\rangle\otimes U \as\ .}
Here the information from internal degrees of freedom imprints on modes $q_1', q_2^{\prime\prime}$ that do not otherwise have large excitation in the Hawking state.
This model does not saturate subadditivity and so is not simple subsystem transfer.  It can however be thought of as a combination of subsystem transfer of the information of two qubits, followed by Hawking pair production.  In this sense, it is similar to \maxsuper.
This model can be described by saying that Hawking production behaves normally, but there is an additional flux of information (hence energy) from the interior of the black hole.  In the limit of a large black hole and slow evaporation rates, this evolution can still be rather innocuous, and not introduce large stresses near the horizon.  A necessary condition for unitary evolution ending with a pure exterior state is that the information transfer rate exceed the rate of new entanglement being created by the Hawking pairs.

These models merely serve as particular examples; as noted they can be generalized to more realistic multi-mode models\refs{\SGunit}, and in the absence of further constraints, evolution could even include both.  We next turn to further comments on general features of black hole evolution.

\subsec{Scrambling and transfer}

Section 3.4 makes the general distinction between information scrambling and transfer in the context of interacting subsystems; let us consider their roles when a black hole interacts with its environment.  Note that one characteristic of the two types of evolution is the timescale on which they operate.  To illustrate this, let us compare various semiclassical predictions with the unitary models that we have described.  

In the semiclassical description of black hole evolution first given by Hawking, the transfer time is effectively infinite:  the information never transfers to the external state (though the calculation certainly fails once the black hole reaches Planck size).  The scrambling time, however, appears gauge-dependent, in accord with the general discussion of sec.~3.4.  Specifically, if we base our description on a set of  ``nice" spatial slices, which are chosen to avoid the strong curvature region (for more details see \SGunit), the excitations have frozen time evolution on the slice and in particular never scramble.  On the other hand, if we use a ``natural" slicing\refs{\NLvC,\SGunit}, such as described by observations of a collection of  satellites freely falling into the black hole, semiclassical evolution of inside particles terminates on timescales $\sim R$ where they encounter strong curvature.  It is not unreasonable to assume that degrees of freedom then scramble, in the absence of a concrete description.  These nice and natural slicings are expected to be related by a unitary transformation -- modulo details of Planck scale dynamics.

For the Page dynamics summarized above, the scrambling time is short, as is the transfer time.  Namely, Page assumes the action of a general random unitary on the internal state, and transfer that begins immediately.  However, as Page shows, the amount of information that is transferred out is very small until the black hole and exterior subsystems are of comparable size.

In the models described in sec.~3.3, information transfer from internal degrees of freedom is immediate.  However, this does not mean that a given bit that has fallen in (or is paired with an outgoing Hawking quantum) immediately begins to transfer.  At one extreme, consider  \evoltwo\ where $\uhat$ is simply nice-slice evolution of LQFT.  A given bit then freezes, until it hits the leftmost position in the state, and is transferred according to \evoltwo.  If there are $\calo(S_{BH})$ total bits, this can take a time $\sim R S_{BH}$.  Similar considerations hold for \evolthree.

Alternately, $\uhat$ could describe more rapid scrambling, resulting in more rapid transfer of a given bit.\foot{After a long enough time, information of a given bit can be recovered on the scrambling time scale \refs{\HaPr}.}  If one only had the picture motivated by natural slices, one might in fact conjecture rapid scrambling.  But, if the nice slice picture is valid, it suggests that there is a gauge where the scrambling is slow.  While one might consider transfer acting on any of the bits, generalizing \evoltwo\ or \evolthree, the picture where they only transfer after a long time, when they have reached the ``leftmost" position, is in a sense ``closest" to the vanishing transfer and scrambling of the semiclassical nice-slice picture.  Indeed, this can be motivated by noting that there are arguments\refs{\QBHB, \NLvC} that the perturbative nice-slice state fails to describe the black hole quantum state after a time $\sim R S_{BH}$.  But, 
one can also consider an intermediate continuum of more rapid scrambling and transfer times in investigating models for the true non-perturbative dynamics.

\subsec{The question of saturation}

In describing information transfer from a black hole to its surroundings, a first question to answer is how close the transfer is to the saturation of subadditivity, described in section 3.2. As shown there, saturation implies that the information transfer is simple subsystem transfer, essentially direct transfer of degrees of freedom (or quanta), whereas departure from this would indicate transfer involving more complicated interactions.  We have noted that either kind of transfer can be described; the former was assumed in \Page, but more detailed models are given for both saturating and non-saturating evolution in \refs{\SGmodels,\SGunit}.

There are motivations for expecting that the information transfer is near saturation.  One reason for this is that, as noted in section 3.2, departure from saturation involves extra excitation.
If we imagine that information transfer from a black hole is a small correction to semiclassical evolution, due to a weak effect, this suggests it involves minimal extra excitation.

A second argument arises from the discussion of section 4.5 of \SGunit, and from the discussion of section 3.  Suppose that the information transfer from the black hole to surroundings is only via couplings to $\hnear$, that it takes the form
\eqn\initwithnear{ {\frac{1}{\sqrt{|A|}}} \sum_{i} \ket{i_C}\ket{i_A} \otimes \ket{\phi_{near}} \otimes \ket{\phi_{far}}\rightarrow  {\frac{1}{\sqrt{|A|}}} \sum_{i} \ket{i_C} \otimes U(\ket{i_A}\ket{\phi_{near}}) \otimes \ket{\phi_{far}}\ ,}
and that subsequently information transfers unitarily to $\hfar$ {\it e.g.} through evolution of LQFT form.  
If, as we have discussed, the relevant modes of $\hnear$  span a space with a relatively small dimension, as summarized in section 2.3, this limits the departure from saturation.  One can think of this limitation as arising from the limited ``bandwidth" of communication through $\hnear$ to the rest of $\hext$.  Specifically, the constraint of small $|\hnear|$ combined with \satbound\ limits the deviation from saturation at each step of the evolution.   Essentially, information transfer to the environment only results from interactions with the BH atmosphere, and restricting the relevant modes of the latter limits the transfer and its deviation from minimality.  

Note that saturation of subadditivity is closely connected with the usual thermodynamic condition of statistical independence of subsystems; in particular, for vanishing mutual information, $S_{AB}=S_A+S_B$.  For a hot body that radiates subsystems (photons, {\it etc.}), one typically assumes such independence.

Also, we saw in \evolthree\ that deviation from saturation can produce extra energy flux.  Indeed, recall that in general $S_A+S_B\geq S_{AB}={\rm const.}$, with equality corresponding to saturation.  So, deviation from saturation increases in a process where $\Delta(S_A + S_B)>0$.  If energy is conserved, this corresponds to
\eqn\Eineq{{dE\over dS_B}< -{dE\over dS_A}\ .}
Specifically, if the energy per bit of excitation of $B$ is $\sim \beta^{-1}\sim 1/R$, then $dE/dS_A \roughly< -\beta^{-1}$.  If so, the black hole can radiate all of its energy before $S_A$ goes to zero, returning us to the paradoxes of remnants or information loss.  The only obvious way to avoid this is if in the non-saturating case, the typical excitation energies of $B$ quanta are lower than $\sim \beta^{-1}$.

\bigskip\bigskip\centerline{{\bf Acknowledgments}}\nobreak

We would like to thank J. Hartle, M. Hastings, D. Marolf, and W. van Dam for useful conversations, and M. Hastings for comments on an earlier version.  This work  was supported in part by the Department of Energy under Contract DE-FG02-91ER40618 and by grant FQXi-RFP3-1008 from the Foundational Questions Institute (FQXi)/Silicon Valley Community Foundation.  Part of this work was carried out while SBG was participating in the KITP program ``Bits, branes, and black holes," and was supported in part by the National Science Foundation under Grant No. NSF PHY11-25915.

\appendix{A}{No information escape via paired states}

In this appendix, we provide the proof of the statement, given in section 2.4, that information cannot escape the black hole if corrections take the form of paired Hawking-like states, \asdf\ -- even if such corrections are large.

This result follows from strong subadditivity \ssa, which can be written equivalently\subadfact
\eqn\ssan{S_{AB}+S_{BC}\geq S_A+S_C\ .}
Setting A = {$\chi_{\text{old}}$} (preexisting state outside black hole), B = {$\chi_{\text{new}}$} (new outgoing particle(s)), and C = {$\hat\psi_{\text{new}}$} (new inside particle(s)), gives
\eqn\deltaS{ \eqalign{
	S(\chi_{\text{old}}\cup\chi_{\text{new}}) + S(\text{pair}) & \geq S(\chi_{\text{old}}) + S(\hat\psi_{\text{new}})\ ,\ {\rm or}, \\
	S(\chi_{\text{all}}) & \geq S(\chi_{\text{old}}) + \left[S(\hat\psi_{\text{new}}) -S(\text{pair})\right]\ .
}}
Looking back at \state, we see that the density matrix for the pair is
\eqn\pairstate{ \eqalign{
	\rho_{\text{pair}} =
		\left( \matrix{
			\bra{\Lambda^0}\ket{\Lambda^0} & \bra{\Lambda^0}\ket{\Lambda^1} & \cdots\\
			\bra{\Lambda^1}\ket{\Lambda^0} & \bra{\Lambda^1}\ket{\Lambda^1} & \cdots\\
			\vdots & \vdots & \ddots
		} \right)
}\ .}
Looking at \pairstate, we see that the density matrix for the new inside state is
\eqn\instatenew{ \eqalign{
	\rho_{\text{new in}} = 
		\left( \matrix{
			\bra{\Lambda^0}\ket{\Lambda^0} & 0 & 0\\
			0 & \bra{\Lambda^1}\ket{\Lambda^1} & 0\\
			0 & 0 & \ddots
		} \right)
}\ .}

Since the density matrix of the new inside state is just the diagonal of the density matrix of the pair, it does not have a lower entropy.  A proof of this claim is included below.  This implies that the square bracket in \deltaS\ is bounded from below by zero.  Our desired result is then
\eqn\Sinc{ S(\chi_{\text{all}}) \geq S(\chi_{\text{old}})\ . }
Therefore, as claimed, the entropy of the black hole cannot decrease.

The necessary claim follows from Klein's inequality.
As a preliminary,  let $\rho$ be a density matrix, and $\sigma$ be its diagonal.
Then $Tr(\rho\ln\sigma) = Tr(\sigma\ln\sigma)$, as trivially follows from diagonality of $\sigma$.
Klein's inequality states
\eqn\klein{Tr(\rho\ln\rho - \rho\ln\sigma) \geq 0\ .}
So, combined with the preceding result, this implies that $S(\sigma)\geq S(\rho)$.

\appendix{B}{Canonical form of a unitary with maximal departure from saturation} 

Ref.~\refs{\Hayden}  proved that all tripartite states that saturate strong subadditivity \ssa\ with equality
\eqn\ssa{S_{AB} + S_{BC} - S_{ABC} - S_B = 0}
have the following structure: $\calh_B$ can be decomposed  $\calh_B = \bigoplus_j \calh_{L_j} \otimes \calh_{R_j}$ and
\eqn\ssastruc{ \rho_{ABC} = \bigoplus_j q_j \rho_{AL_j} \otimes \rho_{R_jC}}
From this the following useful lemma can be proved.

{\bf Lemma:} If $\rho_{ABC}$ is both pure and saturates strong subadditivity, then $\calh_B$ can be decomposed $\calh_B =  \calh_{L} \otimes \calh_{R}$ such that $\rho_{ABC} = \rho_{AL} \otimes \rho_{RC}$; furthermore, $\rho_{AL}$ and $\rho_{RB}$ are both pure.

This is easy to see since purity of $\rho_{ABC}$ and concavity of entropy implies that there is no sum over $j$.  The final clause follows from additivity of entropy.

This lemma can be used to prove the canonical form \maxsuper\ for a unitary with maximal departure from saturation.  The lemma implies that for the resulting state, $\hb$ can be decomposed as $\calh_{L} \otimes \calh_{R}$ such that $\rho_{ABC} = \rho_{AL} \otimes \rho_{CR}$.  Each of these factors are in turn pure, so we have $\rho_{ABC} = \ket{\psi_{AL}}\bra{\psi_{AL}} \otimes \ket{\psi_{RC}}\bra{\psi_{RC}}$.  So far, the unitary has the following structure:
\eqn\maxsuperunit{ U: \frac{1}{\sqrt{|A|}} \sum_i \ket{i_A}\ket{i_C}\ket{\phi_B} \rightarrow \ket{\psi_{AL}} \otimes \ket{\psi_{RC}} \ .}
$C$ is still maximally entangled, and maximally entangled bipartite states are unique up to a choice of basis, so $\ket{\psi_{RC}} = \frac{1}{\sqrt{|A|}} \sum_i \ket{i_R}\ket{i_C}$.  This suffices to prove the aforementioned canonical form \maxsuper.  As a final observation, there is still residual entanglement between $A$ and $B$ determined by $\rho_{AL}$, but independent of the initial state on $A$.

\listrefs
\end